%% file: main.tex
\pdfoutput=1
\documentclass[12pt,a4paper]{article}

\usepackage{ifthen} 
\newboolean{pdflatex}
\setboolean{pdflatex}{true} 

\newboolean{articletitles}
\setboolean{articletitles}{true} 

\newboolean{uprightparticles}
\setboolean{uprightparticles}{false} 

\def\paperauthors{LHCb collaboration} 
\def\paperasciititle{Graph Clustering} 
\def\papertitle{Graph Clustering: a graph-based clustering algorithm for the electromagnetic calorimeter in LHCb} 
\def\paperkeywords{{Calorimeter}, {Graphs}, {Optimization}, {Computing}, {Reconstruction}} 
\def\papercopyright{\the\year\ CERN for the benefit of the LHCb collaboration} 
\def\paperlicence{CC BY 4.0 licence}
\def\paperlicenceurl{https://creativecommons.org/licenses/by/4.0/}

\input{preamble}
\usepackage{longtable} 

\usepackage{algpseudocode}
\usepackage{algorithm}
\usepackage{amsmath}
\usepackage{lineno}

\algnewcommand\algorithmicforeach{\textbf{for each}}
\algdef{S}[FOR]{ForEach}[1]{\algorithmicforeach\ #1\ \algorithmicdo}
\algnewcommand\And{\textbf{and}}

\begin{document}

\renewcommand{\thefootnote}{\fnsymbol{footnote}}
\setcounter{footnote}{1}

\input{title-LHCb-PAPER}


\renewcommand{\thefootnote}{\arabic{footnote}}
\setcounter{footnote}{0}

\cleardoublepage


\pagestyle{plain} 
\setcounter{page}{1}
\pagenumbering{arabic}


\input{body}

\input{acknowledgements}




\addcontentsline{toc}{section}{References}
\bibliographystyle{LHCb}
\bibliography{main,standard,LHCb-PAPER,LHCb-CONF,LHCb-DP,LHCb-TDR}


\end{document}

%% file: preamble.tex
\usepackage[top=1in, bottom=1.25in, left=1in, right=1in]{geometry}

\usepackage{microtype}
\usepackage{lineno}  
\usepackage{xspace} 
\usepackage{caption} 

\usepackage{graphicx}  
\usepackage{color}
\usepackage{colortbl}
\graphicspath{{./figs/}} 

\usepackage{amsmath} 
\usepackage{amssymb}
\usepackage{amsfonts}
\usepackage{upgreek} 

\newcommand*\patchAmsMathEnvironmentForLineno[1]{%
\expandafter\let\csname old#1\expandafter\endcsname\csname #1\endcsname
\expandafter\let\csname oldend#1\expandafter\endcsname\csname
end#1\endcsname
 \renewenvironment{#1}%
   {\linenomath\csname old#1\endcsname}%
   {\csname oldend#1\endcsname\endlinenomath}%
}
\newcommand*\patchBothAmsMathEnvironmentsForLineno[1]{%
  \patchAmsMathEnvironmentForLineno{#1}%
  \patchAmsMathEnvironmentForLineno{#1*}%
}
\AtBeginDocument{%
\patchBothAmsMathEnvironmentsForLineno{equation}%
\patchBothAmsMathEnvironmentsForLineno{align}%
\patchBothAmsMathEnvironmentsForLineno{flalign}%
\patchBothAmsMathEnvironmentsForLineno{alignat}%
\patchBothAmsMathEnvironmentsForLineno{gather}%
\patchBothAmsMathEnvironmentsForLineno{multline}%
\patchBothAmsMathEnvironmentsForLineno{eqnarray}%
}

\usepackage{hyperxmp}

\usepackage[pdftex,
            pdfauthor={\paperauthors},
            pdftitle={\paperasciititle},
            pdfkeywords={\paperkeywords},
            pdfcopyright={Copyright (C) \papercopyright},
            pdflicenseurl={\paperlicenceurl}]{hyperref}

\usepackage[colorinlistoftodos,textsize=scriptsize]{todonotes}

\usepackage[bottom,flushmargin,hang,multiple]{footmisc}

\usepackage[all]{hypcap} 

\input{lhcb-symbols-def} 

\usepackage{cite} 
\usepackage{mciteplus}

%% file: lhcb-symbols-def.tex
\usepackage{xspace} 
\usepackage{upgreek}

\def\MagUp {\mbox{\em Mag\kern -0.05em Up}\xspace}

\ifthenelse{\boolean{uprightparticles}}%
{

 \def\PDelta      {\ensuremath{\Delta}\xspace}                 
 \def\PXi         {\ensuremath{\Xi}\xspace}                 
 \def\PLambda     {\ensuremath{\Lambda}\xspace}                 
 \def\PSigma      {\ensuremath{\Sigma}\xspace}                 
 \def\POmega      {\ensuremath{\Omega}\xspace}                 
 \def\PUpsilon    {\ensuremath{\Upsilon}\xspace}
 \let\oldPi\Pi
 \def\PPi         {\ensuremath{\oldPi}\xspace}

 \def\PB      {\ensuremath{\mathrm{B}}\xspace}                 
                  
 \def\PD      {\ensuremath{\mathrm{D}}\xspace}

 \def\PK      {\ensuremath{\mathrm{K}}\xspace}

 \def\Pi      {\ensuremath{\mathrm{i}}\xspace}

 \def\Ps      {\ensuremath{\mathrm{s}}\xspace}

 \def\thebaroffset{0.0em}
}
{

 \mathchardef\PDelta="7101
 \mathchardef\PXi="7104
 \mathchardef\PLambda="7103
 \mathchardef\PSigma="7106
 \mathchardef\POmega="710A
 \mathchardef\PUpsilon="7107
 \mathchardef\PPi="7105
                  
 \def\PB      {\ensuremath{B}\xspace}                 
                  
 \def\PD      {\ensuremath{D}\xspace}

 \def\PK      {\ensuremath{K}\xspace}

 \def\Pi      {\ensuremath{i}\xspace}

 \def\Ps      {\ensuremath{s}\xspace}

 \def\thebaroffset{0.18em}
}
\newcommand{\offsetoverline}[2][\thebaroffset]{\kern #1\overline{\kern -#1 #2}}%

\makeatletter
\ifcase \@ptsize \relax
  \newcommand{\miniscule}{\@setfontsize\miniscule{4}{5}}
\or
  \newcommand{\miniscule}{\@setfontsize\miniscule{5}{6}}
\or
  \newcommand{\miniscule}{\@setfontsize\miniscule{5}{6}}
\fi
\makeatother

\DeclareRobustCommand{\optbar}[1]{\shortstack{{\miniscule (\rule[.5ex]{1.25em}{.18mm})}
  \\ [-.7ex] $#1$}}

\def\squark    {{\ensuremath{\Ps}}\xspace}

\def\KorKbar {\kern \thebaroffset\optbar{\kern -\thebaroffset \PK}{}\xspace}

\def\D       {{\ensuremath{\PD}}\xspace}

\def\DorDbar {\kern \thebaroffset\optbar{\kern -\thebaroffset \PD}\xspace}

\def\Dp      {{\ensuremath{\D^+}}\xspace}
\def\Dm      {{\ensuremath{\D^-}}\xspace}

\def\DpDm    {\ensuremath{\Dp {\kern -0.16em \Dm}}\xspace}

\def\B       {{\ensuremath{\PB}}\xspace}

\def\BorBbar {\kern \thebaroffset\optbar{\kern -\thebaroffset \PB}\xspace}

\def\Bd      {{\ensuremath{\B^0}}\xspace}

\def\BdorBdbar {\kern \thebaroffset\optbar{\kern -\thebaroffset \Bd}\xspace}

\def\Bs      {{\ensuremath{\B^0_\squark}}\xspace}

\def\BsorBsbar {\kern \thebaroffset\optbar{\kern -\thebaroffset \Bs}\xspace}

\def\Y#1S{\ensuremath{\PUpsilon{(#1S)}}\xspace}

\def\LorLbar     {\kern \thebaroffset\optbar{\kern -\thebaroffset \PLambda}\xspace}

\def\AT#1     {\ensuremath{A_{\mathrm{T}}^{#1}}\xspace}           

\def\C#1      {\ensuremath{\mathcal{C}_{#1}}\xspace}                       
\def\Cp#1     {\ensuremath{\mathcal{C}_{#1}^{'}}\xspace}                    
\def\Ceff#1   {\ensuremath{\mathcal{C}_{#1}^{\mathrm{(eff)}}}\xspace}        
\def\Cpeff#1  {\ensuremath{\mathcal{C}_{#1}^{'\mathrm{(eff)}}}\xspace}       
\def\Ope#1    {\ensuremath{\mathcal{O}_{#1}}\xspace}                       
\def\Opep#1   {\ensuremath{\mathcal{O}_{#1}^{'}}\xspace}                    

\newcommand{\aunit}[1]{\ensuremath{\text{\,#1}}}       

\newcommand{\tev}{\aunit{Te\kern -0.1em V}\xspace}
\newcommand{\gev}{\aunit{Ge\kern -0.1em V}\xspace}
\newcommand{\mev}{\aunit{Me\kern -0.1em V}\xspace}
\newcommand{\kev}{\aunit{ke\kern -0.1em V}\xspace}
\newcommand{\ev}{\aunit{e\kern -0.1em V}\xspace}
 
\newcommand{\mevc}{\ensuremath{\aunit{Me\kern -0.1em V\!/}c}\xspace}
\newcommand{\gevc}{\ensuremath{\aunit{Ge\kern -0.1em V\!/}c}\xspace}
\newcommand{\mevcc}{\ensuremath{\aunit{Me\kern -0.1em V\!/}c^2}\xspace}
\newcommand{\gevcc}{\ensuremath{\aunit{Ge\kern -0.1em V\!/}c^2}\xspace}

\def\gsim{{~\raise.15em\hbox{$>$}\kern-.85em
          \lower.35em\hbox{$\sim$}~}\xspace}
\def\lsim{{~\raise.15em\hbox{$<$}\kern-.85em
          \lower.35em\hbox{$\sim$}~}\xspace}

\def\tell1  {TELL1\xspace}
\def\ukl1   {UKL1\xspace}



%% file: title-LHCb-PAPER.tex

\begin{titlepage}
\pagenumbering{roman}

\vspace*{-1.5cm}
\centerline{\large EUROPEAN ORGANIZATION FOR NUCLEAR RESEARCH (CERN)}
\vspace*{1.5cm}
\noindent
\begin{tabular*}{\linewidth}{lc@{\extracolsep{\fill}}r@{\extracolsep{0pt}}}
\ifthenelse{\boolean{pdflatex}}
{\vspace*{-1.5cm}\mbox{\!\!\!\includegraphics[width=.14\textwidth]{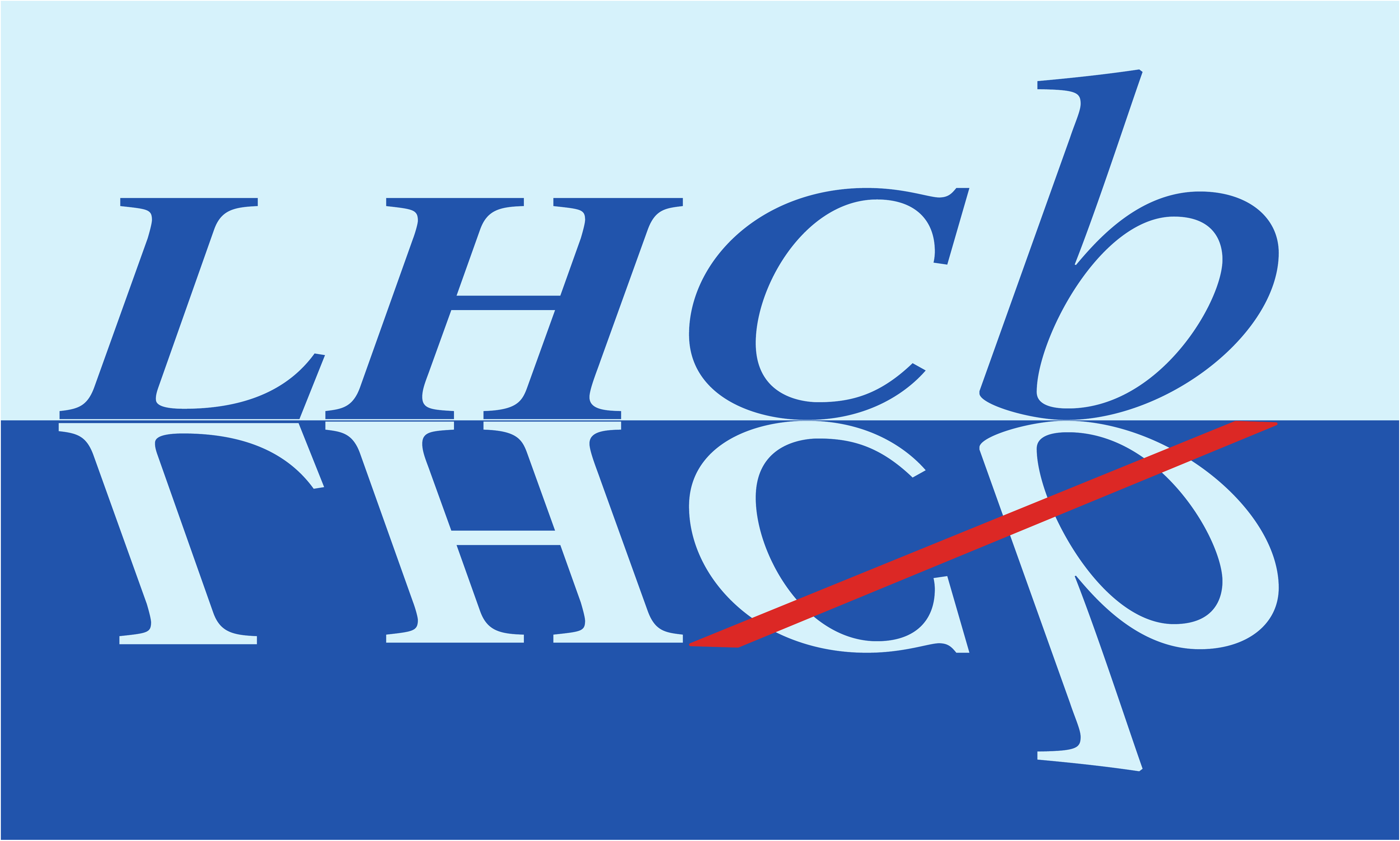}} & &}%
{\vspace*{-1.2cm}\mbox{\!\!\!\includegraphics[width=.12\textwidth]{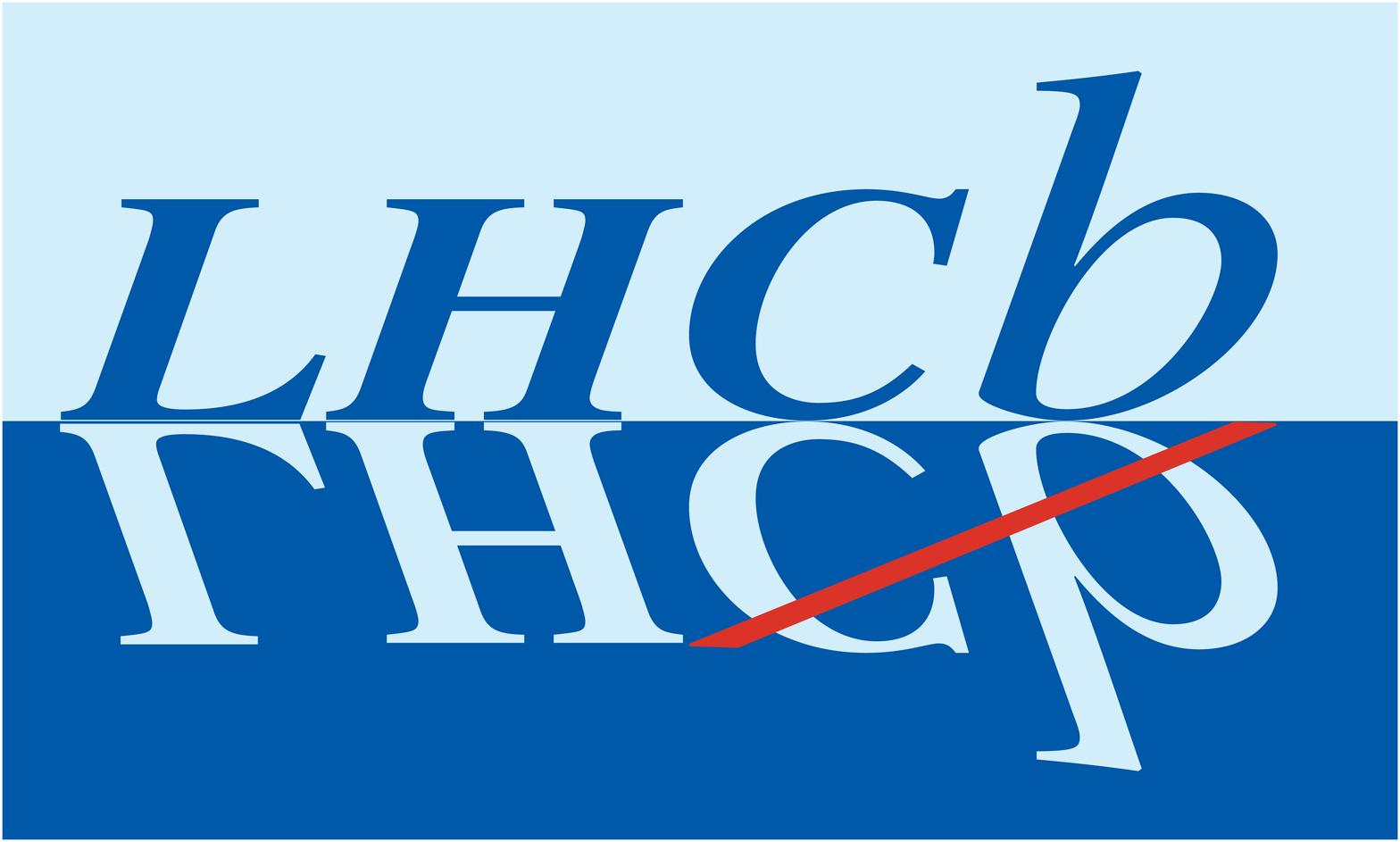}} & &}%
\\
 & & LHCb-DP-2022-003 \\  
 & & \today \\ 
 & & \\
\end{tabular*}

\vspace*{4.0cm}

{\normalfont\bfseries\boldmath\huge
\begin{center}
  \papertitle 
\end{center}
}

\vspace*{2.0cm}

\begin{center}
N. Valls Canudas$^{1}$, M. Calvo Gómez$^{1}$, X. Vilasís-Cardona$^{1}$, E. Golobardes Ribé$^{1}$
\bigskip\\
{\normalfont\itshape\footnotesize

$ ^{1}$Smart Society Research Group, Engineering Department, La Salle Universitat Ramon Llull, Sant Joan de la Salle 42, 08022 Barcelona, Spain
\\
}
\end{center}

\vspace*{0.5cm}

\begin{abstract}
  \noindent
  The recent upgrade of the LHCb experiment pushes data processing rates up to 40 Tbit/s. Out of the whole reconstruction sequence, one of the most time consuming algorithms is the calorimeter reconstruction. It aims at performing a clustering of the readout cells from the detector that belong to the same particle in order to measure its energy and position. This article presents a new algorithm for the calorimeter reconstruction that makes use of graph data structures to optimise the clustering process, that will be denoted Graph Clustering. It outperforms the previously used method by $65.4\%$ in terms of computational time on average, with an equivalent efficiency and resolution. The implementation of the Graph Clustering method is detailed in this article, together with its performance results inside the LHCb framework using simulation data.  
  
\end{abstract}

\vspace*{2.0cm}

\begin{center}
  Submitted to
  Eur.~Phys.~J.~C 
\end{center}

\vspace{\fill}

{\footnotesize 
\centerline{\copyright~\papercopyright. \href{\paperlicenceurl}{\paperlicence}.}}
\vspace*{2mm}

\end{titlepage}


\newpage
\setcounter{page}{2}
\mbox{~}
%
%
%
%

%% file: body.tex
\section{Introduction}
\label{sec1}

LHCb is one of the four main experiments at the LHC at CERN. It consists of a forward-arm spectrometer detector designed to measure the production and decay properties of charm and beauty hadrons with high precision \cite{alves2008lhcb,lhcbColRef}. Starting in 2022, a major upgrade has taken place in order to adapt the luminosity rates of the experiment to the LHC conditions in Run 3. It implies an increment of the instantaneous luminosity by a factor five to $\mathcal{L}=2\times 10^{33} \rm{cm}^{-2}\rm{s}^{-1}$ \cite{RefUG1} and a readout rate of 30MHz, or a maximum data rate of 40 Tbit/s for all the subdetectors. In these conditions, collisions which are of interest for many LHCb analyses reach the MHz level in the detector's geometrical acceptance \cite{Fitzpatrick:1670985}. Therefore, the event selection process is expected to provide an offline-quality reconstruction within the throughput requirements. With the vision of future upgrades implying even tighter time constraints, the LHCb reconstruction needs to be as optimal as possible.

Among the five most time-consuming algorithms in the High Level Trigger 2 (HLT2) sequence of LHCb, the calorimeter reconstruction was the fourth one with the order of $15\%$ of the total cost. To make a significant improvement, the data reconstruction process from this specific sub-detector has been a target to optimise. In this article, a new algorithm for calorimeter reconstruction called Graph Clustering is presented. In terms of execution time, Graph Clustering outperforms the previous method by up to $65.4\%$ with an equivalent efficiency. Overall, it provides an average throughput reduction of $9.8\%$ in the whole HLT2. Furthermore, it is currently the default solution for calorimeter reconstruction in the upcoming Run~3.

This article is aimed to give a background in other reconstruction methodologies used for similar problems, specifically in High Energy Physics, in Section \ref{sec2}. In Section \ref{calo}, an introduction to the Electromagnetic Calorimeter (ECAL) of LHCb is given. Section \ref{sec3} provides an extensive detail of the Graph Clustering implementation.  Finally, in Section \ref{sec4} a review of the performance of the algorithm is given, followed by a discussion and conclusions.

\section{Background}
\label{sec2}

Calorimeter data reconstruction can be understood as a clustering problem, as it aims to group the energy deposits from particles following a set of rules. Classical unsupervised clustering algorithms use extensive recursive functions to create clusters according to metrics related to distance or density \cite{han2011data}. Despite the cluster concept, the calorimeter reconstruction strategy for LHCb has not much in common with classical clustering algorithms, due to the strong physics and execution time requirements.

Focusing on the field of calorimetry in High Energy Physics, the Cellular Automaton has been a benchmark solution for many years \cite{RefCA}. LHCb has been using this strategy for Runs 1 and 2. In 2004 an approach using spanning trees was proposed, using this flexible data structures to exploit the neighbourhood definitions in general calorimeter data \cite{mavromanolakis2004calorimeter}, but it does not consider the cluster separation needed in LHCb. Graph data structures started to appear in the field with the increasing popularity of deep learning in the form of a neural model based on graphs \cite{scarselli2008graph}. Several approaches have used these graph neural networks on layered calorimeters \cite{ju2020graph,qasim2021multi} showing promising results on clustering energy deposits in consecutive calorimeter layers. However, the LHCb calorimeter geometry is bi-dimensional. Within this context, other approaches have also used graph neural networks \cite{qasim2019learning} and convolutional neural networks \cite{RefJoao,valls2021use} with similar conditions as ECAL in LHCb. That said, the inference of some deep learning models is still not mature enough to be incorporated in the LHCb software framework.

Graph structures have demonstrated to be suited for calorimeter data. Hence, the Graph Clustering algorithm stores the calorimeter digits into graphs and makes use of its flexible neighbourhood properties to define the clusters. Moreover, it follows the same reconstruction principles from the Cellular Automaton strategy, which has proved to give a good performance in terms of reconstruction efficiency.

\subsection{Detail of the Electromagnetic Calorimeter}
\label{calo}

\begin{figure}[h]
\centering
\includegraphics[width=0.45\textwidth]{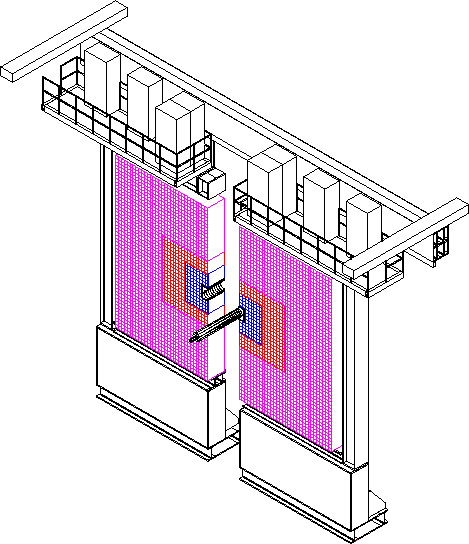}
\caption{The electromagnetic calorimeter 3d view from behind the detector towards the interaction point \cite{RefCalo}.}
\label{fig_ECAL}
\end{figure}

The LHCb experiment has a subset of eight dedicated detectors to acquire data from the particles generated in the LHC collisions. The electromagnetic calorimeter is one of them. Its main purpose is the identification of hadrons, electrons and photons, and the measurement of their energies and positions \cite{RefCalo}. The ECAL has a rectangular shape of $7.8\times 6.3$ $\rm{m}^2$ and is placed perpendicular to the accelerator beam pipe. The energy measurement area is segmented into individual square-shaped modules. Each module is made from lead absorber plates interspaced with scintillator tiles as active material. The general structure is segmented in three different rectangular shaped regions, as can be seen in Figure~\ref{fig_ECAL}.

\vspace{-0.5cm}

Although all modules have the same size of $12\times12$ $\rm{cm}^2$, the number of readout cells on a module depends on the region. The inner region is the closest to the beam pipe and has the highest occupancy of incident particles. Thus, it has the highest granularity among the three regions, with nine readout cells of $4\times4$ $\rm{cm}^2$ per module. The middle region surrounds the inner one and has four readout cells of $6\times6$ $\rm{cm}^2$ per module. The outer region has a single readout cell of $12\times12$ $\rm{cm}^2$ per module.

The output data obtained from the ECAL modules are the values from each readout cell concerning the accumulated energy deposited by incident particles in a collision event with 12 bit precision. Since particles may deposit energy in more than one readout cell, the energy deposits need to be reconstructed and clustered together with the ones belonging to the same particle. This process is precisely the cluster reconstruction for the calorimeter. The current definition of a calorimeter cluster stands for a $3\times3$ block of readout cells around an energy peak. Studies have been done regarding the cluster shapes \cite{abba2013lhcb} where a combination of $2\times2$ and swiss-cross cluster shapes show promising performance for high luminosity, although the $3\times3$ cluster is used as a base for masking other shapes on clusters. Hence, the definition of $3\times3$ readout cell clusters is maintained through all the regions of the detector.

\section{Method}
\label{sec3}

\begin{figure*}[h]
\centering
\includegraphics[width=0.8\textwidth]{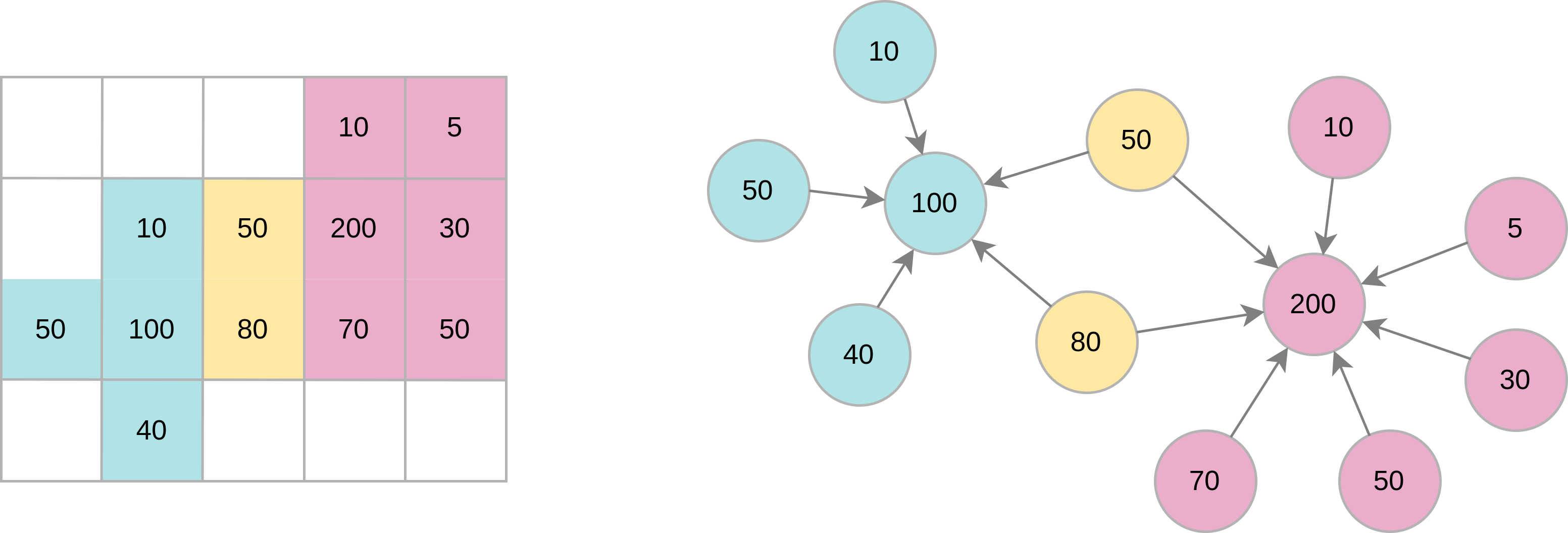}
\caption{An example of two clusters with overlapping cells on the calorimeter on the left and its graph representation on the right.}\label{fig-graph-1}
\end{figure*}

The baseline idea behind the Graph Clustering algorithm is to use graphs as a data structure to store the event digits. It transforms the calorimeter digits into independent graph structures, where only relevant digits for a cluster are contained into isolated graphs. Following graph theory nomenclature, each energy digit from an event is represented as a vertex $v$ in the graph, also called node. The relations between digits, representing links to the same cluster, are defined as directional edges $(u, v)$ between the source digit node $u$ and the target node $v$. By design, the target nodes of all edges in the graph are the seeds of the reconstructed clusters, where a seed is defined as a local maximum energy digit in the calorimeter grid over a threshold of $50$ MeV in transverse energy. With this, the cluster seeds can be easily identified as nodes with only incoming edges. Furthermore, a node can be linked to more than one seed if it is susceptible to have energy deposits from more than one particle. These particular cases are called overlap cells. Overall, the graph derived from an event may contain structures like the example shown in Figure \ref{fig-graph-1}.

The following subsections describe in detail the four steps needed in the Graph Clustering reconstruction process.

\subsection{Sorting}

To achieve the mentioned representation of the digits, the algorithm needs to make an efficient insertion of the edges into the graph structure. Since all the edges are based on the cluster seeds, the initial key point is to identify seed candidates. As defined previously, a cell in the ECAL grid is considered a seed if it is a local maximum and has a minimum transverse energy value of 50 MeV. A local maximum in this context defines a cell that has the highest energy value among its distance one neighbours in the calorimeter grid. This definition is the same as the one used in the Cellular Automaton algorithm \cite{RefCA}.

In order to process the seed candidates of an event in the first place, all the digits above 50 MeV need to be sorted by decreasing traverse energy value. In the proposed algorithm, the sorting is computed using Introspective Sorting \cite{musser1997introspective}, which is a hybrid sorting algorithm that combines three different methods to provide fast average performance and optimal worst-case performance.

\subsection{Insertion}

The role of the insertion step is to build the graph edges between the event digits such that the graph structures of Figure \ref{fig-graph-1} are obtained. A pseudo-code notation of this process is stated in Algorithm \ref{insertion}.

\begin{algorithm}
\caption{Graph insertion}\label{insertion}
\begin{algorithmic}[1]
\small{
\State $G \Leftarrow $ directional weighted graph
\ForEach{$energy, id \in sortedDigits$}
    \If{$id$ not inserted in $G$}
        \If{$id$ is local maxima}
            \State add node $id$ in $G$
            \ForEach{$n_{energy}, n_{id} \in $ neighbours of $id$}
                \State add node $ne_{id}$ and edge $(ne_{id}, id, w=1)$ in $G$
                \If{$id$ is a merged $\pi^0$ candidate}
                    \State add $id$ and $n_{id}$ to $mergedPi0$
                \EndIf
            \EndFor
        \EndIf
    \ElsIf{$id \in mergedPi0$}
        \State $seed = $ first seed from $id$ in $G$
        \ForEach{$n_{energy}, n_{id} \in $ neighbours of $id$}
            \If{$energy > n_{energy} \And n_{id}$ not in $G$}
                \State add node $ne_{id}$ and edge $(ne_{id}, id, w=1)$ in $G$
            \EndIf
        \EndFor
    \EndIf
\EndFor
}
\end{algorithmic}
\end{algorithm}

It essentially iterates over each sorted digit. That digit may have already been inserted in the graph. If so, this means it is a neighbour of a more energetic digit. In that case, it cannot be a seed since there cannot be two adjacent maxima by construction, except for the case of merged $\pi^0$s, which is explained in section \ref{mergedpi0}. Therefore, that digit is not inserted. On the other hand, if the digit has not yet been inserted on the graph, it can be either a seed or a residual digit, meaning it is not a local maximum and does not have any seed on its neighbourhood. To distinguish between the two, the algorithm checks if that digit is a local maximum. If it is the case, the seed is inserted in the graph together with all its neighbour digits linking them with edges to the seed. The default weight value for all edges is one.

Additionally, if a merged $\pi^0$ candidate is identified following the metrics described in section \ref{mergedpi0}, there is a second seed added to the same cluster. The neighbours of the second seed are also linked with an edge to the first seed if they are not already inserted and if its energy deposit is lower than the energy of the first seed.

At the end of the insertion step, the clusters are already grouped in the graph. However, the overlap cases still need to be processed to adjust the weight of the overlap edges.

\subsubsection{Merged $\pi^0$ case}
\label{mergedpi0}

\begin{figure*}[h]
\centering
\includegraphics[width=0.80\textwidth]{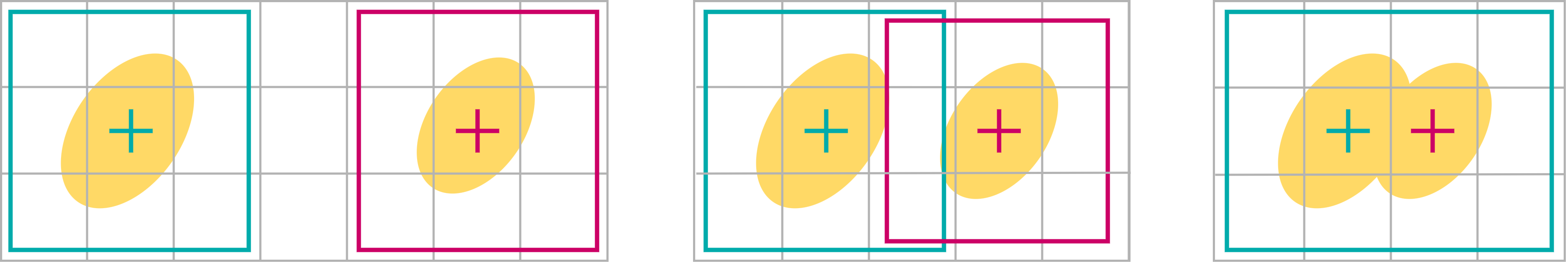}
\caption{Diagram representation of $\pi^0$ cluster cases on the calorimeter. From left to right: the two photons are separable and without overlap, it is a resolved $\pi^0$. The two photons are separable but have three overlapping digits, it is however a resolved $\pi^0$. The two photons are not separable, it is a merged $\pi^0$ and is reconstructed as a single cluster bigger than $3\times 3$.}\label{fig-pi0-diagram}
\end{figure*}

One of the reconstruction requirements for the LHCb calorimeter is the correct identification of neutral pions, $\pi^0$, which decay into two photons before reaching the calorimeter. Depending on the energy and momentum of the $\pi^0$, the two photons arrive at the calorimeter with a certain separation. If the photons are distanced enough to be reconstructed as two separate clusters, it is called a resolved $\pi^0$. Otherwise, the two photons may travel very close to each other and reach the calorimeter at one cell distance or less. In that case, the reconstruction is done as a single cluster, since the definition of maxima does not allow two adjacent cluster seeds. When photons are not separable, it is then called merged $\pi^0$ case. Hence, the super-cluster from a merged $\pi^0$ can be bigger than the $3\times 3$ window around the seed as can be seen in Figure \ref{fig-pi0-diagram}.

\begin{figure}[h]
\centering
\includegraphics[width=0.59\textwidth]{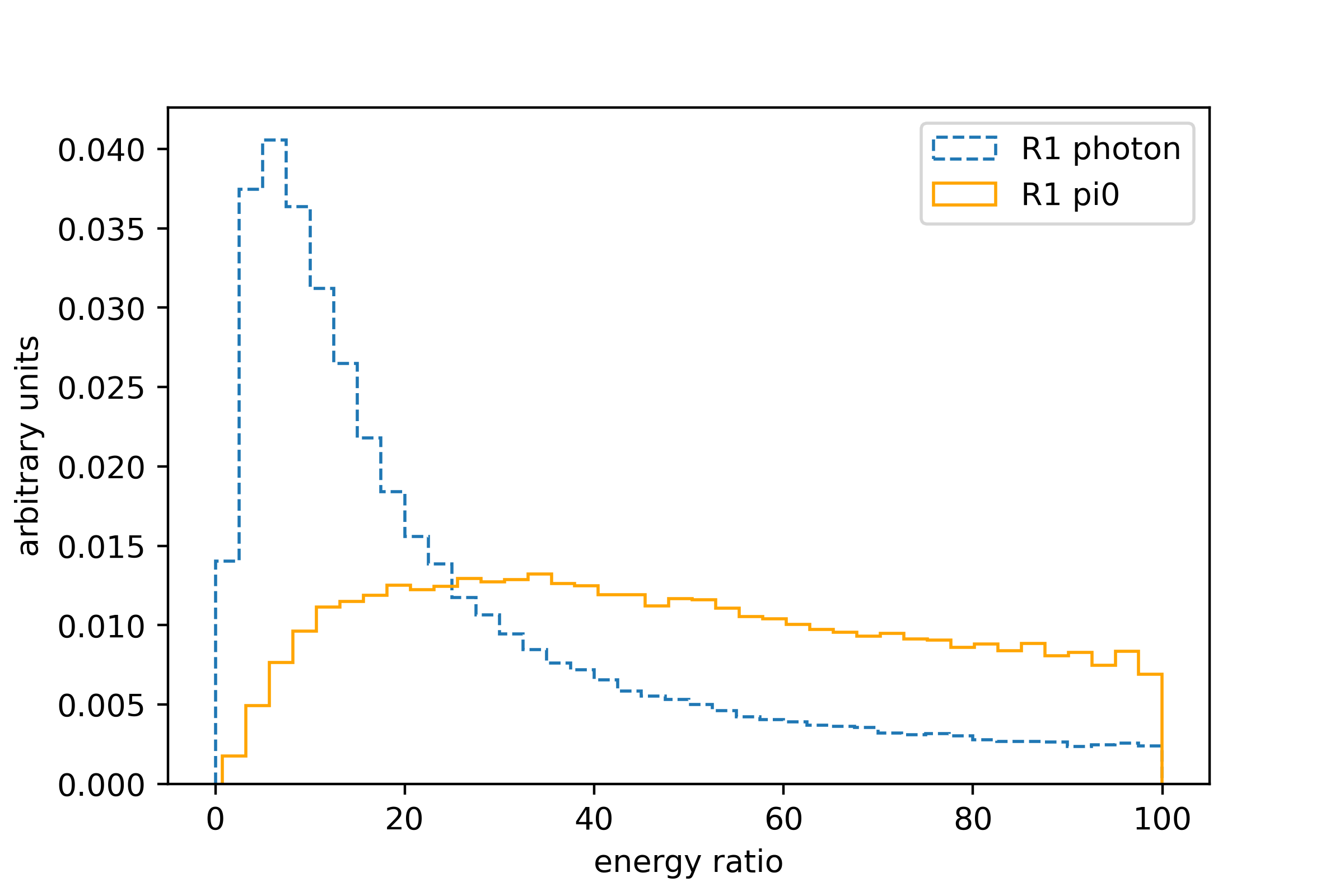}
\caption{Normalized histogram of the energy ratio between the second most energetic digit and the cluster seed for photon samples and  $\pi^0$ samples.}\label{fig-energy-ratio}
\end{figure}

The cluster shape used in $\pi^0$ reconstruction tools in LHCb is a mask of $5\times 5$ cells around the seed \cite{deschamps2003photon}. Therefore, the residual energy outside the $3\times 3$ window of a merged $\pi^0$ is crucial, as it contains part of the energy from the second photon. That is why the Graph Clustering algorithm adapts the shape of potential merged $\pi^0$ candidates, expanding the cluster up to the neighbours of the second most energetic digit in the cluster. The only source of information available to make this selection at run time is the energy deposits of the $3\times 3$ cluster. Therefore, we have studied the relation between the two most energetic digits as a ratio labeled $R1$ to define a threshold on which to make the cluster expansion. Over 46.000 samples of single $\pi^0$ deposits from $B^0 \rightarrow \pi^+\pi^-\pi^0$ decays simulated using Run 3 conditions have been studied. Figure \ref{fig-energy-ratio} shows a normalised histogram of the $R1$ ratio for $\pi^0$ samples and also for photon samples from $B^0 \rightarrow K^*\gamma$ in Run 3 conditions. Given the difference in the two distributions, the threshold for $R1$ is set to 25. The chosen value ensures that the residual energy left outside the cluster is less than $9\%$ for the studied $\pi^0$ samples and that the cluster expansion affects an average of $8.2\%$ of the clusters in an event. Further studies have determined that small variations around $10\%$ of the selected threshold value do not significantly change the time complexity of the algorithm nor the $\pi^0$ resolution. Moreover, a second threshold for merged $\pi^0$ candidates concerning the minimum energy of the cluster seed is set to $1000$ MeV, as only high energy $\pi^0$s are reconstructed as merged according to the data studied. 

\subsection{Connected Components}

All the clusters are already grouped together after the insertion step. To retrieve them from the graph, we need to search for the sub-graphs where all the nodes are connected to each other by some path, ignoring the direction of edges. This is defined by the weakly connected components of the original graph. In the proposed algorithm, this is implemented as a depth-first search \cite{ginsberg2012essentials}, which explores an entire graph exploring all its branches as far as possible before backtracking. Its time complexity is $O(\mid V\mid+\mid E\mid)$ \cite{cormen2022introduction} where $V$ is the number of vertices or nodes in the graph and $E$ is the number of edges. Once all the vertices of the graph are visited, the groups of nodes on each weakly connected component are obtained.

\subsection{Analysis of Clusters}
\label{analysis}

The processing of each weakly connected component can be done independently of the others, since it contains only the relevant nodes for a cluster. In case there is more than one cluster in a connected component, the overlap fractions are calculated and assigned to the respective edge weights. Either way, the independent clusters are then stored as reconstructed clusters. Algorithm \ref{analysis_alg} shows a pseudo-code of this step of the reconstruction. Only connected components with more than one vertex are considered as reconstructed clusters. Any isolated vertex is likely to be residual energy deposits from a cluster and should not be considered a reconstructed cluster itself.

\begin{algorithm}
\caption{Analysis of connected components}\label{analysis_alg}
\begin{algorithmic}[1]
\small{
\ForEach{$wcc \in weaklyConnectedComponents$}
    \If{$wcc.size() > 1$}
        \State calculate overlap weights (Algorithm \ref{overlap-weights})
        \ForEach{$id \in wcc$}
            \If{$id$ in-edges $> 1 \And id$ out-edges $== 0$}
                \State add $id$ as a cluster seed to $clusters$
                \ForEach{$vertex$ connected to $id$}
                    \State add $vertex$ as entry of $id$ in $clusters$
                \EndFor
            \EndIf
        \EndFor
    \EndIf
\EndFor
}
\end{algorithmic}
\end{algorithm}

In Algorithm \ref{overlap-weights} a pseudo-code version of the overlap weight calculation is provided. This function iterates through all the vertices in a connected component. It searches for overlap vertices, identified by having two or more output edges, and accumulates the energy of all the connected nodes on all the clusters involved in the overlap, including the energy of the overlapping node equally fractioned for every involved cluster. Then, a weight is computed for every overlapping edge as the fraction between the energy of the target cluster and the sum of all the clusters involved in the overlap.

\begin{algorithm}
\caption{Calculate overlap weights}\label{overlap-weights}
\begin{algorithmic}[1]
\small{
\State $clusterEnergy \Leftarrow $ empty map
\ForEach{$vertex \in wcc$}
    \If{$vertex$ out-edges $>= 2$}
        \ForEach{$end\_vertex \in vertex$ out-edges}
            \State $energy =$ accumulate energy from the nodes linked to $end\_vertex$.
            \State $energy +=$ $end\_vertex$ energy / num out-edges.
            \State store $energy$ to $clusterEnergy$
        \EndFor
        \State $totalEnergy = $ accumulate $clusterEnergy$ energies with entries $\in vertex$ out-edges
        \ForEach{$end\_vertex \in vertex$ out edges}
            \State $weight = \frac{clusterEnergy \text{ at } end\_vertex}{totalEnergy}$
            \State set edge $(vertex, end\_vertex, w=weight)$
        \EndFor
     \EndIf
\EndFor
}
\end{algorithmic}
\end{algorithm}

At the end of this last step, all the reconstructed clusters from an event are stored.

\section{Results}
\label{sec4}

All the algorithm tests have been done within the GAUDI framework \cite{TheLHCbCollaboration:2310827,GAUDI}. For comparison purposes, this paper evaluates the performance of the Graph Clustering algorithm and the Cellular Automaton algorithm as it has been a benchmark solution until now. Both are tested with the same Monte Carlo data from $B^0 \rightarrow K^*\gamma$ simulations using Run~3 conditions.

The quality of the reconstruction in calorimeter algorithms in LHCb is evaluated using metrics of efficiency, energy resolution and position resolution. The efficiency is defined as the fraction between reconstructed particles over reconstructible particles in a set of events. Reconstructible particles are photons that have deposited at least $90\%$ of its energy in the calorimeter cells. On the other hand, reconstructed particles are reconstructible particles matching a cluster from which at least $90\%$ of its energy belongs to that particle. This ratio is later referred to as match fraction. Table \ref{tab1} shows that Graph Clustering has a higher efficiency than the Cellular Automaton, with $1.02\%$ more reconstructed clusters.

\begin{table*}[h]
\begin{center}
\caption{Efficiency results in number of reconstructed versus reconstructible clusters from 80,000 $B^0 \rightarrow K^*\gamma$ events.}\label{tab1}%
\begin{tabular}{@{}llll@{}}
\hline\noalign{\smallskip}
Algorithm & Reconstructible & Reconstructed & Efficiency ($\%$)\\
\noalign{\smallskip}\hline\noalign{\smallskip}
Graph Clustering & 43234   & 35313  & $81.68\pm0.19$  \\
Cellular Automaton & 43234   & 34872  & $80.66\pm0.19$  \\
\noalign{\smallskip}\hline
\end{tabular}
\end{center}
\end{table*}

On the other hand, the cluster resolution metric aims to measure the difference in energy and position between the reconstructed clusters and the associated Monte Carlo particles. Resolutions are evaluated for $\gamma$ and $\pi^0$ particles. For both cases, we evaluate the difference in position on the $X$ and $Y$ axis and the difference in energy as a percentage. For $\gamma$ resolution, a total of $80.000$ simulation samples of $B^0\rightarrow K^*\gamma$ decays have been used, and another $80.000$ samples of $B^0 \rightarrow \pi^+\pi^-\pi^0$ decays have been used for $\pi^0$ resolution. The study accounts for all the clusters with a match fraction higher than $0.9$ since it is the standard match threshold for a cluster to be considered reconstructed in terms of efficiency.

\begin{figure}[h]
\centering
\includegraphics[width=0.49\textwidth]{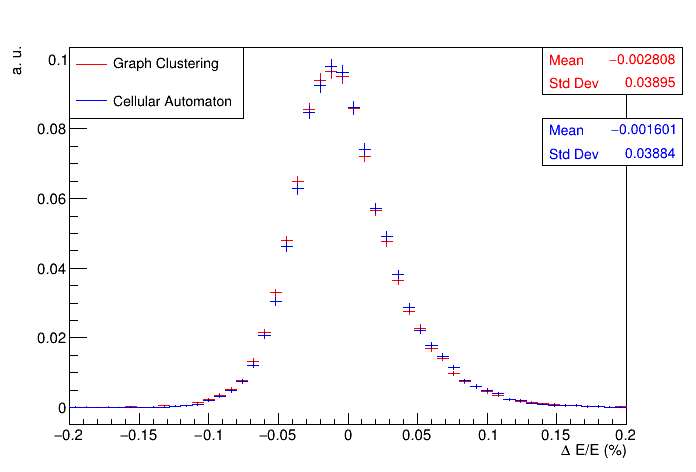}
\includegraphics[width=0.49\textwidth]{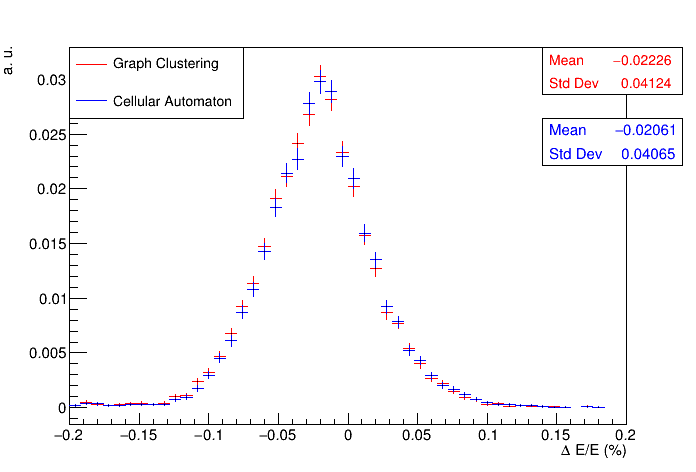}
\caption{Normalized histograms of the energy resolution with no corrections for clusters with a match fraction over $0.9$ using $\gamma$ samples in the top plot and $\pi^0$ samples in the bottom plot.}\label{res-energy}
\end{figure}

Figure \ref{res-energy} shows the energy distribution for both methods, before any corrections are applied \cite{corrections}, where $\Delta E$ stands for the difference in reconstructed energy and truth energy of a cluster. It can be seen that for both $\gamma$ and $\pi^0$ samples the two distributions look very alike. For energy resolution, Graph Clustering is slightly more shifted to negative values, but overall it can be said that the resolution in energy is equivalent to the Cellular Automaton one.

\begin{figure}[h]
\centering
\includegraphics[width=0.49\textwidth]{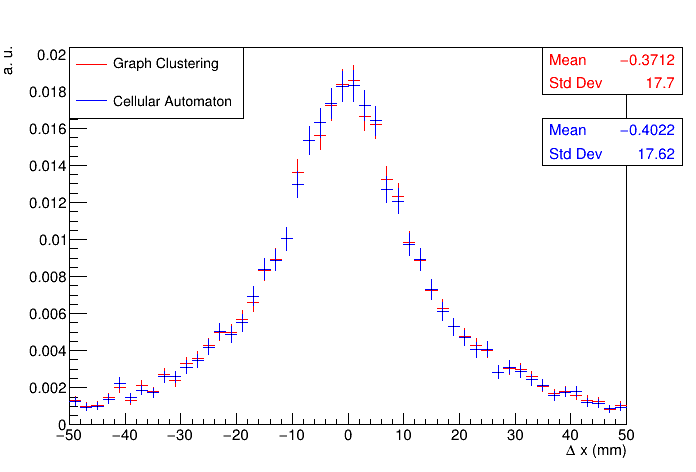}
\includegraphics[width=0.49\textwidth]{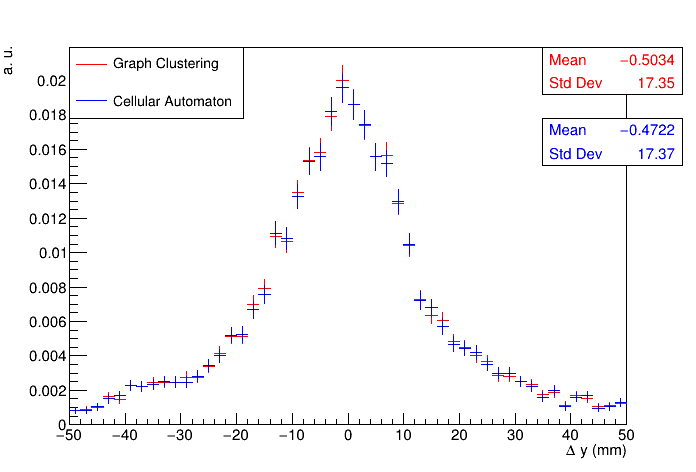}
\caption{Normalized histograms of the X axis resolution at the top and the Y axis resolution at the bottom. Both using $\pi^0$ samples and clusters with a match fraction over $0.9$ with no corrections.}\label{res-pos}
\end{figure}

Regarding the position resolution, Figure \ref{res-pos} shows that the $x$ and $y$ distributions have again an equivalent behavior for both methods. For simplicity, only the $\pi^0$ resolutions are shown for position, since the differences with $\gamma$ samples are minimal.

\begin{figure}[h]
\centering
\includegraphics[width=0.6\textwidth]{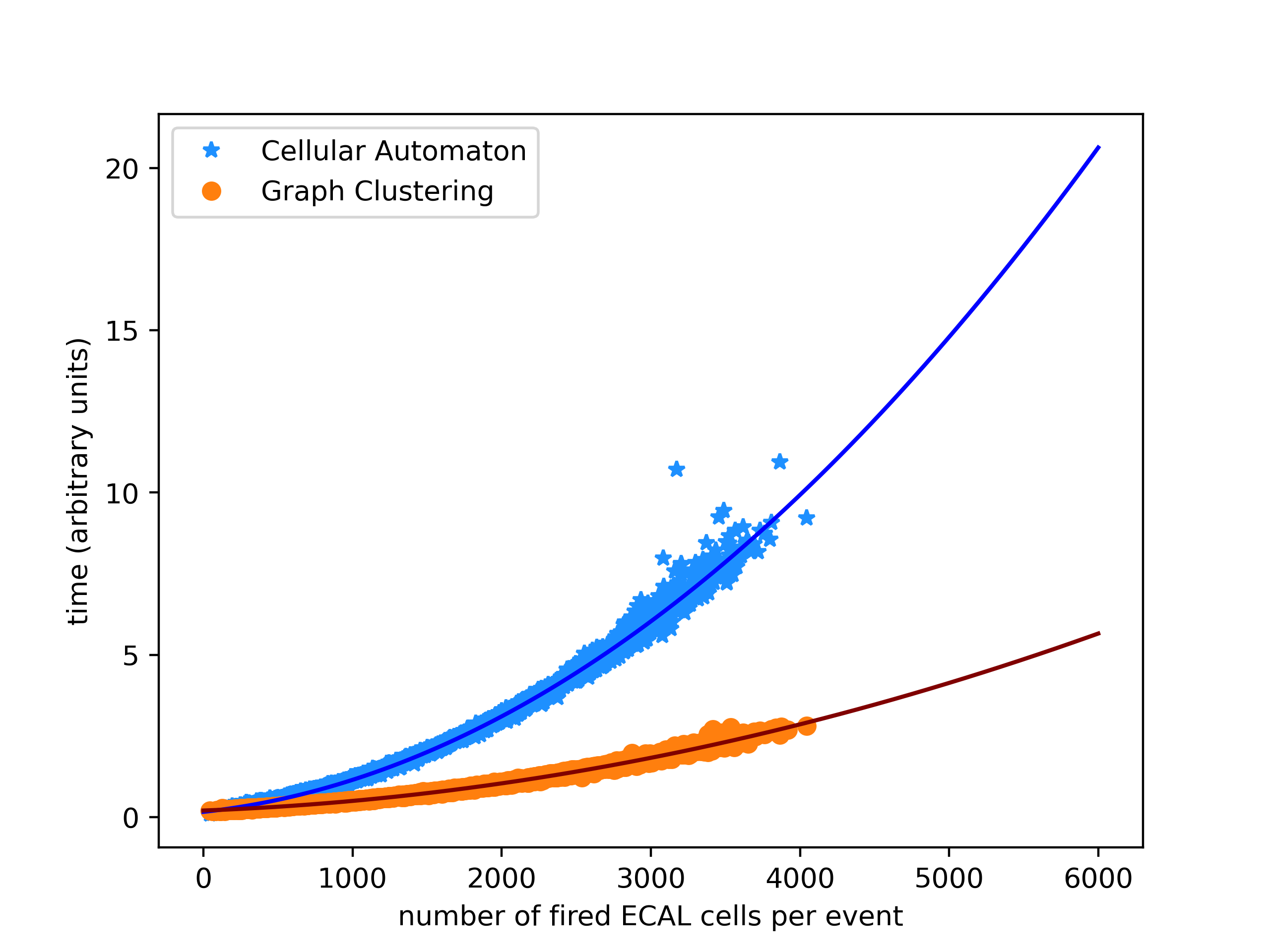}
\caption{Execution time measured in arbitrary units as a function of the number of digits per event for the Cellular Automaton algorithm and the Graph Clustering algorithm. On top of them, a fitted curve for every algorithm is shown.}\label{fig-time}
\end{figure}

Regarding the execution time, it is defined as the time elapsed between the first and the last lines executed in an algorithm. Figure \ref{fig-time} shows a plot of the execution time in arbitrary units as a function of the number of digits per event. The plotted time measurements are obtained as the average measured time from all the events with the same number of digits, from a total of $100.000$ events from $B^0\rightarrow K^* \gamma$ simulation. As can be seen from the figure, for events with less than 150 digits, the Cellular Automaton is faster. However, from that point on, Graph Clustering outstands the benchmark algorithm showing a flatter complexity curve. Furthermore, the average number of digits per event from the analysed samples is 1520 digits. At that complexity level, Graph Clustering is $65.4\%$ faster than Cellular Automaton on average.

\section{Discussion and Conclusions}
\label{sec5}

Graph Clustering has shown to improve the computational complexity of the calorimeter reconstruction in LHCb. Furthermore, it is the default reconstruction solution for the ongoing Run 3 data taking period. The baseline of the algorithm is to reproduce the same reconstruction steps as in the previously used algorithm, the Cellular Automaton, but with an optimized codification using graph data structures. Hence, it is expected and observed to have similar results compared to the benchmark in terms of efficiency and resolution.

Graphs have demonstrated to be suited for calorimeter reconstruction. Within the proposed implementation, such data structures also provide a flexible interpretation of the neighbour cells in the calorimeter grid. This could also be used to adapt the shape of the clusters to an optimized pattern depending on the region at reconstruction time and significantly accelerate its execution. Currently, the definition of an optimal cluster shape for ECAL clusters is being studied considering pileup and overlap effects as well as precision. 

Within the steps of the presented Graph Clustering, as mentioned in section \ref{analysis}, the analysis of each connected component is completely independent of the rest of the graph. Although it is not the most time consuming part of the algorithm, it represents a $27.3\%$ of the total algorithm's execution time, which could benefit from parallel execution. In the context of the first level of the trigger system (HLT1) ran in GPUs, calorimeter reconstruction is at a preliminary stage. The current implementation builds simplified clusters with lower efficiency and resolution than the benchmark. In that direction, there is currently work in progress on adapting the presented Graph Clustering logic to a CUDA algorithm optimized for parallel architectures.

As a final conclusion, the complexity curve that Graph Clustering exhibits makes it a useful alternative for other calorimeters with higher occupancy. Furthermore, the vision of future upgrades in the LHCb calorimeter is a challenging opportunity to test the limits of this algorithm.

%% file: acknowledgements.tex
\section*{Acknowledgements}
%
%

The authors would like to thank the LHCb computing and simulation teams for their support and for producing the simulated LHCb samples used in the paper. Specially the RTA team for their help with code optimization and the integration into the LHCb framework. This research was funded by Ministerio de Ciencia e Innovación grant number~PID2019-106448GB-C32.